\documentclass[twocolumn,showpacs,preprintnumbers,amsmath,amssymb]{revtex4}

\usepackage{graphicx}
\usepackage{dcolumn}
\usepackage{bm}

\begin{document}

\preprint{APS/123-QED}

\title{Mean first passage time for nuclear fission and
the emission of light particles}

\author{Helmut Hofmann$^{1}$}
\author{Fedor A. Ivanyuk$^{1,2}$}
\affiliation{1) Physik Department, TUM, D-85747 Garching, Germany\\
2) Institute for Nuclear Research, 03028 Kiev, Ukraine}

\date{11.04.03}

\begin{abstract}
The concept of a mean first passage time
is used to study the time lapse over which a fissioning system may
emit light particles. The influence of the "transient" and "saddle
to scission times" on this emission are critically examined. It is
argued that within the limits of Kramers' picture of fission no
enhancement over that given by his rate formula need to be
considered.
\end{abstract}

\maketitle

{\em Introduction ---}\label{intro}
 Fission at finite thermal excitation is characterized by the
evaporation of light particles and $\gamma$'s. Any description of
such a process must rely on statistical concepts, both with
respect to fission itself as well as with respect to particle
emission. For decades it has been customary to describe
experiments in terms of particle \cite{FN-nat-par}  and fission
widths, where the former, $\Gamma_{\rm{n}}$, is identified through
the evaporation rate and the latter $\Gamma_{\rm{f}}$ is given by
the Bohr-Wheeler formula $\Gamma_{\rm{f}}\equiv \Gamma_{\rm{BW}}$
for the fission rate. Often in the literature this is referred to
as the "statistical model". It was only in the 80's that
discrepancies of this procedure with experimental evidence was
encountered: Sizably more neutrons were seen to accompany fission
events than given by the ratio $\Gamma_{\rm{n}}/\Gamma_{\rm{BW}}$
(for a review see e.g. \cite{pauthoe}). A possible enhancement of
that ratio is found if the fission width $\Gamma_{\rm{BW}}$ is
replaced by the $\Gamma_{\rm{K}}$ of Kramers \cite{kram}. In this
seminal paper he pointed to the deficiency of the picture of Bohr
and Wheeler in that it discards the influence of couplings of the
fission mode to the nucleonic degrees of freedom. Such couplings
will in general reduce the flux across the barrier, mainly because
of the reduction of the energy in the fission degree of freedom
$Q$ which may then fall below the barrier.
This $Q$ is meant to represent the most likely path in a
multidimensional landscape of shape degrees of freedom.

In Kramers' picture this effect is realized through the presence
of frictional and fluctuating forces (intimately connected to each
other by the fluctuation dissipation theorem). Presently it is
understood that Kramers' "high viscosity limit" applies (for a
microscopic justification see \cite{hofrep, hiry}), in which case
the rate formula writes\begin{equation}\label{kram-rate}
\Gamma_{\rm{K}}= \frac{\hbar\varpi_{\rm{a}} }{ 2\pi}
\exp\left(-\frac{E_{\rm{b}}}{T}\right)\,
\left(\sqrt{1+\eta_{\rm{b}}^2} - \eta_{\rm{b}}\right)
=\frac{\hbar}{\tau_{\rm{K}}}
.\end{equation} Here, $T$ and $E_{\rm{b}}$ stand for temperature
and barrier height, $\varpi_{\rm{a}}$ for the frequency of the
motion around the minimum at $Q=Q_{\rm{a}}$ and
$\eta_{\rm{b}}=(\gamma/(2M \varpi))_{\rm{b}}$ for the dissipation
strength at the barrier (at $Q=Q_{\rm{b}}$) with $\gamma$ being
the friction coefficient and $M$ the inertia. For the sake of
simplicity we will assume these coefficients not to vary along the
fission path; otherwise the formula must be modified \cite{hiry}.
For vanishing dissipation strength (\ref{kram-rate}) reduces to
the Bohr-Wheeler formula (simplified to the case that the
equilibrium of the nucleons can be parameterized by a
temperature).

Commonly, formula (\ref{kram-rate}) is derived (see
e.g.\cite{scheuho}) in a time dependent picture solving the
underlying Fokker-Planck equation for special initial conditions
with respect to the time dependence of the distribution function
\cite{FN-Kram-Smol}. Their choice is intimately related to the
picture of a compound reaction, in that the decay process is
assumed to be independent of how the compound nucleus is produced.
The latter in a sense represents a nucleus in a quasi-equilibrium
such that the previous, pre-equilibrium stages need not be
considered explicitly. This assumption is valid as long as the
decay of that system takes longer than the equilibration time. To
some large extent such a situation is indeed given at not too high
excitations, as then the nucleons may stay inside this nuclear
complex for a sufficiently long time. However, the circumstances
are less clear with respect to the collective modes, in particular
to the fission degree of freedom itself --- which for large
damping probably is among the slowest ones present. Whereas the
corresponding
kinetic momentum $P=M\dot Q$
may safely be assumed to
equilibrate sufficiently fast, this may not be so for the
coordinate $Q$.
Thus, assuming the system to be located initially around the
supposedly pronounced "ground state" minimum of the static energy
at $Q=Q_{\rm{a}}$, the initial width in $Q$ may still be at one's
disposal. In its true spirit the compound picture would suggest
taking the equilibrium value, determined by the temperature and,
in harmonic approximation, by the stiffness of the potential.
Often, however, one starts with a sharp distribution of zero
width. In any case, the current across the barrier needs some
finite time to build up. This apparent {\em delay} of fission was
interpreted \cite{graliwei,bhagran} as if there was the {\em
additional} possibility of emitting light particles {\em beyond}
the measure given by $\Gamma_{\rm{n}}/\Gamma_{\rm{K}} \,> \,
\Gamma_{\rm{n}}/\Gamma_{\rm{BW}}$.

If besides collective motion also particle emission is studied
explicitly in a time dependent picture,
as done in the Langevin approach \cite{pobaridi, mactheo}, such an
effect is included automatically. Problems arise, however, if one
tries to imitate this delay in statistical codes which are in use
for analyzing experimental results. Such codes  apply static
probabilities derived in {\em time independent} reaction theory.
It is not obvious how this method may be reconciled with the
picture of fission delay, the "transient effect". In the present
note we like to shed some light on this problem by exploiting the
concept of a mean first passage time (MFPT). Before we shall come
to that we want to examine a little closer the time dependent
case. We will concentrate on over-damped motion, as in this case
the MFPT can be evaluated from an analytic formula. Moreover, for
slow motion the transient time gets larger, such that the feature
we want to discuss becomes even more obvious.

{\em Time dependent current across the barrier
---}\label{cur-crobar} In the time dependent picture just
described the boundary conditions in $Q$ (and $P$ if present) are
chosen to make sure that the distribution vanishes at infinity.
Calculations of the current $j(t)$ across the barrier then
typically imply a behavior
as exhibited in Fig.\ref{ran385}.
In all cases the asymptotic value of $j_{\rm{b}}(t)$ is seen to
follow the law $\Gamma_{\rm{K}} \exp(-\Gamma_{\rm{K}} \,t/\hbar)$,
shown by the fully drawn straight line.
The differences at short times are due to the following different
initial conditions:
\begin{figure}[h]
\begin{center}
\includegraphics[width=0.48\textwidth]{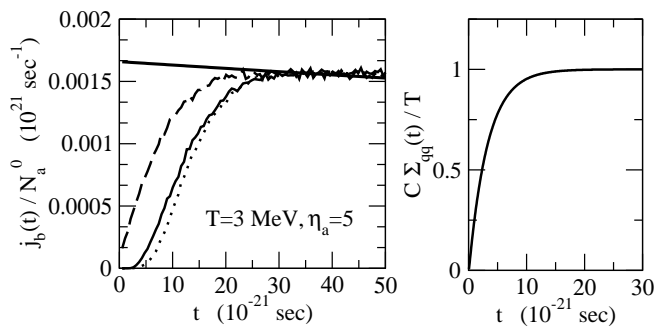}
\vspace{-6mm}
\caption{The current across the barrier for different initial
conditions, see text.} \vspace{-6mm}
\label{ran385}
\end{center}
\end{figure}

(i) For the dashed and dotted curves the system starts out of
equilibrium in $Q$; the dashed curve corresponds to the current at
the barrier $j_{\rm{b}}(t) = j(Q_{\rm{b}},t)$ and the dotted one
to that in the scission region $j_{\rm{sc}}(t) =
j(Q_{\rm{sc}},t)$,
beyond which the fragments separate.
The equilibrium is defined by the oscillator potential by which
the $V(Q)$ around $Q_{\rm{a}}$ may be approximated.

(ii) For the fully drawn line the system starts at $Q_{\rm{a}}$
sharp. The obvious delay by about $5-10 \cdot 10^{-21}$ sec is
essentially due to the relaxation of $Q$ to the quasi-equilibrium
in the well. This feature is demonstrated on the right by the
$t$-dependence of the width in $Q$ (exhibited in terms of
fluctuations of the potential energy $C\Sigma_{qq}=C(\langle
Q^2\rangle - \langle Q\rangle^2)/2$).

\begin{figure}[h]
\begin{center}
\includegraphics[width=0.48\textwidth]{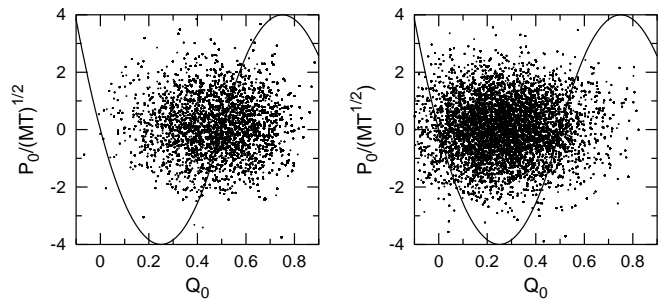}
\vspace{-5mm}
\caption{Samples of initial points which overcome the saddle
within a time $\tau_s$: Left part: $\tau_s\simeq
\tau_{\rm{trans}}$; right part $\tau_s\simeq \tau_{\rm{K}} $.}
\vspace{-6mm} \label{points35}
\end{center}
\end{figure}

The figure clearly demonstrates remarkable uncertainties in the
very concept of the "transient effect". First of all it is seen,
that the "transient" time $\tau_{\rm{trans}}$, defined as the time
the current $j_{\rm{b}}(t)$ needs to reach its asymptotic
behavior, depends strongly on the initial conditions. Moreover,
there is considerable arbitrariness in choosing time zero: If the
calculation were repeated at some later time $t_0
>\tau_{\rm{trans}}$, the same features would be seen! In the end this is due
to the very fact that the whole effect only comes about because in
the initial distribution there are favorable parts for which it is
easiest to reach the barrier. This is demonstrated in
Fig.\ref{points35}. There, those points of the initial equilibrium
are sampled which cross the barrier after some given time
$\tau_s$. On the right a sufficiently large $\tau_s$ was chosen
such that {\em greater parts} of the initial distribution have
"fissioned". As exhibited on the left, for the much shorter time
$\tau_s\simeq \tau_{\rm{trans}} $, only a small fraction of points
have succeeded in doing this, namely those which started close to
the barrier (for under-damped motion also more favorite initial
momenta would play a role, see \cite{HoIv-rauisch,
HoIv-MFPT-big}). The vast majority of particles is still waiting
to complete the same motion but at later times! This aspect is
important, not only for an understanding of the essentials of the
concept of the MFPT \cite{vankampen,gardiner-STM,Risken}, but also
in respect to the evaporation of neutrons. Indeed, even for
$\tau_{\rm{K}}\gtrsim t \gg \tau_{\rm{trans}}$ there is ample time
for them to be emitted from {\em inside} the barrier.

 The calculations have been performed by simulating the Langevin
equations exploiting a locally harmonic approximation similar to
that of \cite{scheuho, hofrep} for Kramers' equation, for the
following parameters: $T=3$ MeV, $E_b=8$ MeV, $\hbar\varpi_a=1$
MeV and $\eta_a=5$. The potential was constructed from two
oscillators, one upright and one upside down, joined with a smooth
first derivative.

\begin{figure}[h]
\begin{center}
\includegraphics[width=0.48\textwidth]{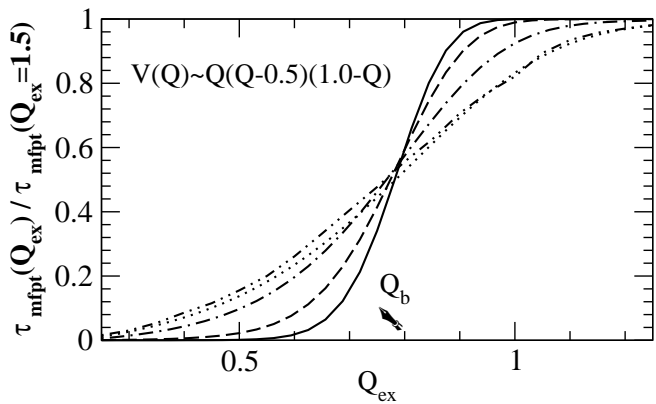}
\vspace{-5mm}
\caption{MFPT for a cubic potential normalized to its asymptotic
value, shown by the solid, dashed, dotted-dashed and dotted curves
which correspond to $T/E_b=0.1, 0.2, 0.5, 1.0$.  The dashed double
dotted curve represents a calculation within the Langevin approach
for $T/E_b=1$.} \vspace{-6mm} \label{cubic32a}
\end{center}
\end{figure}

{\em The mean first passage time ---}\label{conc-MFPT} Within a
Langevin approach the concept of MFPT may be described as follows.
Suppose that at $t=0$ particles start at the potential minimum
$Q_{\rm{a}}$. Because of the fluctuating force there will be
trajectories $i$ which pass a certain exit point $Q_{\rm{ex}}$
first at some time $t_i$, the {\em first passage time}. The {\em
mean}-FPT $\tau_{\rm{mfpt}} (Q_{\rm{a}} \to Q_{\rm{ex}})$ is
defined by the average $\langle t_i \rangle $ over all
possibilities. In order to really obtain the mean {\em first}
passage time the $i$ has to be removed from the ensemble once it
has exited the interval at $Q_{\rm{ex}}$: the "particle" can be
said to be absorbed at $Q_{\rm{ex}}$ (such that one may speak of
an "absorbing barrier"). As the potential $V(Q)$ is assumed to
rise to infinity for $Q\to -\infty$, any motion to the far left
will bounce back: the region $Q\to -\infty$ acts as a "reflecting
barrier". A calculation of the MFPT with the Langevin equation is
shown in Fig.\ref{cubic32a} by the dashed double dotted curve and
seen to be very close to the result obtained by exploiting special
solutions of the Smoluchowski equation, which we want to address
now.

Fortunately, the Smoluchowski  approach allows one to derive an
analytic formula for the $\tau_{\rm{mfpt}}$
\cite{vankampen,gardiner-STM,Risken}. As one knows, the
Smoluchowski equation represents that of Kramers for over-damped
motion. For its solution $K(Q,t\,|\,Q_{\rm{a}},0)$ the initial
condition for the particles to start at $Q_{\rm{a}}$ is given by
$\lim_{t\to 0 }K(Q,t|Q_{\rm{a}},0)= \delta(Q-Q_{\rm{a}})$, which
is identical to the one used for the fully drawn line of
Fig.\ref{ran385}. For constant friction and temperature one gets
\begin{equation}
\label{MFPT-smo} \tau_{\rm{mfpt}}=
\frac{\gamma}{T}\int_{Q_{\rm{a}}}^{Q_{\rm{ex}}}du
\exp\left[\frac{V(u)}{T}\right]\int_{-\infty}^udv
\exp\left[-\frac{V(v)}{T}\right].
\end{equation}
This expression may be derived as follows: The probability of
finding at time $t$ the particle still inside the interval
$(-\infty,~Q_{\rm{ex}})$ is given by
$W(Q_{\rm{a}},t)=\int_{-\infty}^{Q_{\rm{ex}}}\,
 dQ\times$ $K(Q,t\,| \,Q_{\rm{a}},0) $. Hence, the probability for it
to leave the region during the time lapse from $t$ to $t+dt$ is
determined by $-d W = -(\partial W(Q_{\rm{a}},t)/\partial t)\,
dt$, such that the average time becomes
$\tau_{\rm{mfpt}}(Q_{\rm{a}}\to Q_{\rm{ex}})= -\int t \,d W$ which
turns into
\begin{eqnarray}
\label{MFPT-prop}
\tau_{\rm{mfpt}}(Q_{\rm{a}}\to Q_{\rm{ex}})& =&
\int_0^\infty
\int_{-\infty}^{Q_{\rm{ex}}}\,K(Q,t\,|\,Q_{\rm{a}},0) \, dQ \,dt
\nonumber\\&=&
\int_0^\infty dt\,t\, j(Q_{\rm{ex}},t\,|\,Q_{\rm{a}},0) \,.
\end{eqnarray}
These formulas are associated to  the special
boundary conditions with respect to the
 coordinate mentioned before, the reflecting barrier at $Q\to
-\infty$ and an absorbing barrier at $Q_{\rm{ex}}$. In particular
for the latter feature it is not permitted to use in
(\ref{MFPT-prop}) the currents shown in Fig.\ref{ran385}.
Inserting them blindly would indeed lead to expressions for
$\tau_{\rm{mfpt}}$ in which the $\tau_{\rm{trans}}$ appears
\cite{FN-app-curr}. {\em This is in clear distinction to the
correct form (\ref{MFPT-smo}).} Actually, the derivation of
(\ref{MFPT-smo}) involves proper solutions of that equation which
is "adjoint" to the Smoluchowski equation, and which describes
motion backward in time. In Fig.\ref{cubic32a} we show the
dependence of $\tau_{\rm{mfpt}}(Q_{\rm{a}}\to Q_{\rm{ex}})$ on
$Q_{\rm{ex}}$ as given by (\ref{MFPT-smo}) calculated for a cubic
potential. Evidently, the MFPT needed to reach the saddle  at
$Q_{\rm{b}}$ is {\em exactly half the asymptotic value}. The
latter may be identified as the mean fission life time
$\tau_{\rm{f}} \equiv \tau_{\rm{mfpt}}(Q_{\rm{a}}\to Q_{\rm{ex}}
\gg Q_{\rm{b}})$. For the typical conditions under which Kramers'
rate formula (\ref{kram-rate}) is valid for overdamped motion, the
identity of $\tau_{\rm{f}}\equiv \tau_{\rm{K}}$ to the asymptotic
value of the MFPT can be proven analytically \cite{gardiner-STM}.
Another remarkable feature seen in Fig.\ref{cubic32a} is the
insensitivity of the MFPT to the exit point for small and large
$Q_{\rm{ex}}$. Actually, in clear distinction to the transient
time the MFPT is also insensitive to the starting point. This
latter property shall be exhibited in a forthcoming paper
\cite{HoIv-MFPT-big}.

\begin{figure}[h]
\begin{center}
\includegraphics[width=0.48\textwidth]{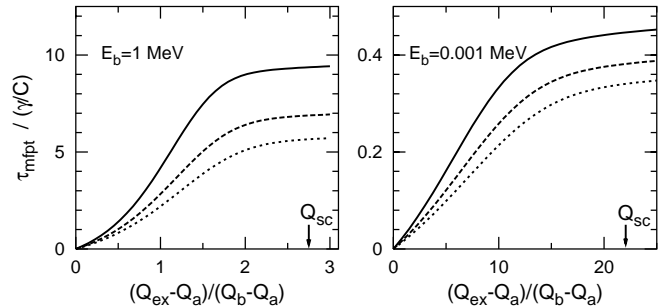}
\caption{The MFPT for the case of small and vanishing barriers for
$T=2,3,4$ MeV from top to bottom. The $Q_{\rm{sc}}$ corresponds to
a scission point 20 MeV below the barrier. } \vspace{-5mm}
\label{smallEb6}
\end{center}
\end{figure}
As a most interesting feature, the MFPT can be calculated also for
cases of small barriers where Kramers' formula does not apply. We
show in Fig.\ref{smallEb6} results of evaluations of formula
(\ref{MFPT-smo}) for $E_b=1$ MeV as well as for a practically
vanishing barrier. It is seen that even in the latter case the
$\tau_{\rm{mfpt}}(Q_{\rm{a}}\to Q_{\rm{ex}}) $ reaches a plateau
for sufficiently large $Q_{\rm{ex}}$. This asymptotic value,
however, is no longer determined by $\tau_{\rm{K}}$. Nevertheless,
the $\tau_{\rm{mfpt}}$ seems to be long enough for neutrons to be
evaporated before scission. This may be seen as follows. The
neutron width typically is of the order of $1\dots 3$ MeV  (for
medium heavy nuclei \cite{pobaridi}). According to \cite{hiry} the
$\gamma/C$ roughly increases linearly in $T$, being about
$1~\hbar$/MeV at $T=2$ MeV and about $3~\hbar$/MeV at $T=3$ MeV.
Taking the values of $\tau_{\rm{mfpt}}$  from the right part of
Fig.\ref{smallEb6} and close to the plateau  one gets a width
$\Gamma_{\rm{mfpt}}=\hbar/\tau_{\rm{mfpt}}$ of the order of $0.9
\dots 2.3 $ MeV, and thus comparable to the neutron width.
Fig.\ref{smallEb6} also shows that a small increase of the barrier
by 1 MeV enlarges the $\tau_{\rm{mfpt}}$ drastically.
{\em Discussion ---}\label{discu}
It should be evident from the previous discussion that in the very
concept of the MFPT there is no room for a transient effect. After
all, formula (\ref{MFPT-smo}) is based on {\em exact solutions of
the transport equation which satisfy the same initial condition as
those used for the plots in Fig.\ref{ran385} --- albeit different
boundary conditions in coordinate space.}
Moreover, as exhibited in Fig.\ref{points35}
the evaluation of the MFPT takes into account an average over {\em
all} initial points, as is warranted by the definition of the MFPT
through the probability distribution $-dW$. Contrasting this
feature, and as outlined in the second section, the transient
effect only represents a minor part of the initial population,
namely that one which reaches the barrier first. Discarding the
rest implies ignoring the many particles which are still moving
inside the barrier for times typically much longer than
$\tau_{\rm{trans}}$. Hence, neutrons from deformations
corresponding to that region may not only be emitted within
$\tau_{\rm{trans}}$ but within $\tau_{\rm{mfpt}}(Q_{\rm{a}}\to
Q_{\rm{b}})$, which turns out to be just half of the total fission
time $\tau_{\rm{K}}$. Of course, this discussion shows that it is
also not correct to argue in favor of "additional" neutrons which
might be emitted within the saddle to scission time
$\tau_{\rm{ssc}}$ introduced in \cite{hofnix}. As one may guess
from Fig.\ref{cubic32a}, like the $\tau_{\rm{trans}}$, the
$\tau_{\rm{ssc}}$ does not appear to be in accord with the MFPT
either: The time the fissioning system stays together is not
determined by motion in the immediate neighborhood of the barrier.
On average it takes half the full decay time to move beyond the $
Q_{\rm{b}}$ to the $ Q_{\rm{ex}}$ at which the $\tau_{\rm{mfpt}}$
reaches its plateau value.

These findings suggest that one simply estimates the emission rate
of neutrons over fission from the ratio $\Gamma_{\rm{n}}/
\Gamma_{\rm{K}}$ --- provided one may trust the potential to be of
the simple form underlying the rate formula (\ref{kram-rate}).
Anything else does not seem to be in accord with an appropriate
application of Kramers' or Smoluchowski's equations. This does not
rule out other, complementary effects which originate in more
complicated situations. For instance, in case that in the scission
region the potential becomes flat again or even develops a minimum
the system is forced to stay there longer than given by the
$\tau_{\rm{K}}$ of eq.(\ref{kram-rate}) --- implying additional
time for evaporating neutrons. Likewise it is conceivable that the
{\em initial} stage of the whole reaction is to be described with
a {\em different transport model}.
Such modifications are already suggested when the average neutron
emission time $\tau_{\rm{n}}$ becomes comparable to or even
smaller than the relaxation time $\tau_{\rm{micro}}$ for the
nucleonic degrees of freedom as a whole. Transport equations are
justified only if this $\tau_{\rm{micro}}$ is the smallest time
scale present, in comparison to both neutron emission as well as
to collective motion. The $\tau_{\rm{micro}}$ turns out to be of
the order of $1-2 \cdot 10^{-22}$ sec, no matter whether it is
estimated within linear response theory with collisional damping
or within a random matrix approach (see \cite{hofrep}). Applying
the Weisskopf estimate for the neutron emission time (or modified
versions of it) \cite{pobaridi} at large temperatures one easily
gets values of $\tau_{\rm{n}}$ of the order of or smaller than
$\tau_{\rm{micro}}$. This reflects a situation of {\em
pre}-equilibrium rather than that assumed in the quasi-static
picture necessary for the application of Fokker-Planck equations.
In conclusion we may say that deviations of experimental results
from the standard value of $\Gamma_{\rm{n}}/\Gamma_{\rm{K}}$ ought
perhaps to be understood as a strong indication of the relevance
of these complementary effects, which unfortunately have for the
most part been unconsidered. This might require one to re-examine
analyzes of experiments which over the past decade or so have
followed the conventional line.

The authors benefitted greatly from a collaboration meeting on
"Fission at finite thermal excitations" in April, 2002, sponsored
by the ECT* ('STATE' contract).  One of us (F.A.I.) would like to
thank the Physik Department of the TUM for the hospitality
extended to him.


\begin{thebibliography}{99}
\bibitem{FN-nat-par} To
 simplify the discussion we will not distinguish the nature of the
 "particles" and in this sense include $\gamma$'s in this notation.
\bibitem{FN-Kram-Smol}For Kramers' equation this involves
the coordinate $Q$ and a momentum $P$. The latter is absent in the
Smoluchowski equation for overdamped motion.
\bibitem{FN-app-curr}Approximating the form of the current by
$j_{\rm{b}}(t) = j_{\rm{norm}}
(1-e^{-\Gamma_{\rm{trans}}t/\hbar})\exp\left(-\Gamma_{\rm{K}}t/\hbar\right)$
one would falsely get $\tau_{\rm{mfpt}}(Q_{\rm{b}}) =
\hbar/(\Gamma_{\rm{K}}+\Gamma_{\rm{trans}}) +
\hbar/\Gamma_{\rm{K}}$.
\end{thebibliography}
\end{document}